\documentclass[12pt]{iopart}

\usepackage{footnote}
\expandafter\let\csname equation*\endcsname\relax
\expandafter\let\csname endequation*\endcsname\relax
\usepackage{graphics} 
\usepackage{graphicx}% Include figure files
\usepackage{dcolumn}% Align table columns on decimal point
\usepackage{bm}% bold math
\usepackage{subfigure}
\usepackage{multirow}
\usepackage{float}
\usepackage{amsmath}
\usepackage{esvect}
\begin{document}

\title{Identified particle spectra in Pb-Pb, Xe-Xe and p-Pb collisions with Tsallis blast-wave model}
\author{Guorong Che, Jinbiao Gu, Wenchao Zhang$^{*}$, Hua Zheng}
\address{School of Physics and Information Technology, Shaanxi Normal University, Xi'an 710119, People's Republic of China}
\eads{wenchao.zhang@snnu.edu.cn}
\begin{abstract}

 \noindent We investigate the identified hadrons transverse momentum ($p_{\rm T}$) spectra  in Pb-Pb (Pb-Pb, Xe-Xe, p-Pb) collisions at $\sqrt{s_{\rm NN}}=$ 2.76 (5.02, 5.44, 5.02) TeV in the framework of Tsallis-blast wave (TBW) model with a linear transverse velocity profile and with a constant velocity profile. In this model, the Tsallis temperature ($T$), the average radial flow velocity ($\langle \beta \rangle$) and the degree of non-equilibrium ($q$) of the system are common for all hadrons when a combined fit is performed to the $p_{\rm T}$ spectra of different particles at a given centrality.  It is found that the model can describe the particle spectra well up to 3 GeV/c. For both profiles, the transverse flow velocity decreases from central to peripheral collisions while the non-extensive parameter $q$ exhibits the opposite behavior, indicating a more rapid expansion and less off-equilibrium of the system in more central collisions. Moreover, we observe that in central collisions $\langle \beta \rangle$ and $q$ ($T$) from the fit with the linear profile are smaller (is slightly larger) than those (that) with the constant profile, while in peripheral collisions $\langle \beta \rangle$, $T$ and $q$ from the former are compatible with those from the latter. We also derived and discussed the relation between the Tsallis temperature and the thermal temperature. In addition,  to check whether a scenario of an early freeze-out of strange particles at the LHC exists, the particle spectra are investigated by grouping them into strange and non-strange hadrons. The combined fit gives an insight on the degree of non-equilibrium, the radial flow and the Tsallis temperature of the system at the kinetic decoupling.  It provides a comparison between the results at different energies in the same collision system and the results in different collision systems at the same or similar energy.
  %In order to check whether there is a dependence of the fit results on the form of the transverse velocity profile, two cases are considered: one is that $\beta$ grows linearly with the the radius ($r$) of the emitting source, the other one is that $\beta$ does not depend on $r$.
%The spectra of deuterons and $^{3}\rm He$ in Pb-Pb (p-Pb) collisions at 2.76 (5.02) TeV are also investigated in the framework of the TBW model. 
\end{abstract}
%\submitto{\jpg}

\maketitle

\section{Introduction}
\label{sec:intro}

The transverse momentum ($p_{\rm T}$) spectra of identified particles are significant observables in high-energy heavy-ion collisions. They can be utilized to investigate the dynamics of particle production. In the low $p_{\rm T}$ region, particle production is governed by soft physics and described by non-perturbative theory or model, such as the Boltzmann-Gibbs blast-wave (BGBW) model\cite{BGBW_dist}.  In the high $p_{\rm T}$ region, it is dominated by hard processes and described by perturbative quantum chromodynamics (pQCD).

The BGBW model has been widely used in the description of the particle spectra in nucleus–nucleus (AA) and proton–nucleus (pA) collisions at the Relativistic Heavy Ion Collider (RHIC) and the Large Hadron Collider (LHC)\cite{STAR_1,STAR_2,STAR_3,LHC_1, LHC_2, LHC_3}. In this model, the shape of the particle spectrum depends on two parameters: the temperature at the kinetic freeze-out $T_{f}$ and the average transverse flow velocity $\langle \beta \rangle$. In refs.\cite{STAR_1,STAR_2,STAR_3}, the STAR collaboration had simultaneously fitted  the $\pi^{\pm}$, $K^{\pm}$ and $p(\bar{p})$ spectra in the low $p_{\rm T}$ region produced in Au-Au collisions at $\sqrt{s_{\rm NN}}=$ 7.7 (11.5, 19.6, 27, 39, 62.4, 130, 200) GeV with this model. They found that at a given collision energy $\langle \beta \rangle$ increased with centrality, indicating more rapid expansion in more central collisions. Moreover, for central collisions, $\langle \beta \rangle$ showed a flat trend for the lowest three energies, and then a steady  increase up to 200 GeV. In refs.\cite{LHC_1, LHC_2, LHC_3}, the ALICE collaboration had investigated these light hadron spectra in Pb-Pb (Pb-Pb, p-Pb) collisions at 2.76 (5.02, 5.02)  TeV. They observed that for central Pb-Pb collisions $\langle \beta \rangle$ was slightly larger at 5.02 TeV than the one at 2.76 TeV. Moreover,  $\langle \beta \rangle$ was smaller in central p-Pb collisions than that in Pb-Pb collisions at 5.02 TeV.

In the BGBW model, there is a strong assumption that the system will reach a local thermal equilibrium at some instant of time and then undergo a hydrodynamic evolution. However, in fact the initial condition for the hydrodynamic evolution fluctuates event by event\cite{initial_fluc}. This fluctuation may leave footprint on the particle spectra in the low and intermediate $p_{\rm T}$ region due to its incomplete wash-out by the subsequent interactions at either the quark-gluon plasma phase or the hadronic phase\cite{TBW_1, initial_fluc_1, initial_fluc_2, initial_fluc_3}. In order to take the effect of the fluctuation into account, in ref.\cite{TBW_1}, the authors had changed the particle emission source distribution from the Boltzmann distribution to the Tsallis distribution\cite{Tsallis_dist}. The Tsallis blast-wave (TBW) model was then used to study the $\pi^{\pm}$, $K^{\pm}$, $p(\bar{p})$, $\phi$, $\Lambda(\bar{\Lambda})$ and $\Xi^{-}(\bar{\Xi}^{+})$ spectra in Au-Au collisions at 200 GeV. They found that $\langle \beta \rangle$ and the Tsallis temperature $T$ increased with centrality while the non-extensive parameter $q$ showed the opposite behavior. In ref.\cite{TBW_2}, this model was extended to the spectra of strange and non-strange hadrons. It was observed that for central collisions strange hadrons have smaller $q$ and  $\langle \beta \rangle$ while higher $T$ than non-strange hadrons, indicating that the formers possibly decouple earlier than the latters.

In ref.\cite{TBW_v2}, the authors had generalized the BGBW model by taking the azimuthal variations in the transverse expansion rapidity and in the density of the source element into account and found that it can successfully reproduce the elliptic flow of $\pi$, $K$ and $p$ in Au-Au collisions at 130 GeV. In this generalized model,  $\beta$ does not depend on the radius ($r$) of the emitting source, while in the original BGBW model, $\beta$ does rely on $r$. In this paper,  as a complementary study to those conducted in refs.\cite{TBW_1,TBW_2}, the TBW model is fitted simultaneously to identified particle spectra at a given centrality in Pb-Pb (Pb-Pb, Xe-Xe, p-Pb) collisions at 2.76 (5.02, 5.44, 5.02) TeV. In order to investigate the dependence of the fit results on the choice of the transverse velocity profile, two cases are considered: one is that $\beta$ relies on $r$ linearly, the other one is that $\beta$ does not depend on $r$. It is found that in both cases the model can describe the particle spectra well up to 3 GeV/c. In addition,  to check whether a scenario of an early freeze-out of strange particles at the LHC exists, the particle spectra are investigated by grouping them into strange and non-strange hadrons. The combined fit gives an insight on the degree of non-equilibrium, the radial flow and the Tsallis temperature of the system at the kinetic decoupling.  It provides a comparison between the results at different energies in the same collision system and the results in different collision systems at the same or similar energy.%Finally, in order to see whether the kinetic freeze-out conditions for light nuclei are identical to those for light hadrons, the investigation is extended to the spectra of deuterons ($d$) and $^{3}\rm He$ in Pb-Pb (p-Pb) collisions at 2.76 (5.02) TeV. 

The organization of this paper is as follows. In section \ref{sec:method}, we briefly describe the TBW model.  In section \ref{sec:results_and_discussions}, the results of the TBW model in Pb-Pb, Xe-Xe and p-Pb collisions  are presented and some discussions are followed. Finally, we give the conclusion in section \ref{sec:conclusions}.

\section{The TBW model}\label{sec:method}

With the TBW model in refs.\cite{TBW_1,TBW_2}, we express the invariant differential yield of identified particles at mid-rapidity as 
\begin{equation}
  \small
  \begin{split}
\frac{d^{2} N}{2\pi p_{\rm T} d p_{\rm T}dy}& \propto m_{\rm T} \int_{-Y}^{+Y} \cosh (y_{s}) d y_{s} \int_{0}^{R}rdr\int_{-\pi}^{+\pi} d \phi_{b}  \int_{-\pi}^{+\pi}d \phi_{p}\\
 &\times \left[1+\frac{q-1}{T} \left(m_{\rm T} \cosh (y_{s}) \cosh(\rho)-p_{\rm T} \sinh (\rho) \cos (\phi_{p}-\phi_{b})\right)\right]^{-1 /(q-1)},\label{eq:eTBW}
\end{split}
\end{equation}
where $y$ and $m_{\rm T}$ are the rapidity and transverse mass of identified particles, $y_{s}$ ($Y$) is the rapidity of the emitting source (beam),  $\phi_{p}$ and $\phi_{b}$  are, respectively, the azimuthal angles of the emitted particle velocity and the flow velocity with respect to the $x$ axis in the reaction plane. The azimuthal direction of the boost, $\phi_{b}$, is deemed as the same as the azimuthal angle of the emitting source in coordinate space, $\phi_{s}$. It is assumed that the kinetic freeze-out temperature $T_{f}$ in the BGBW model fluctuates from event to event. As shown in ref.\cite{q_meaning}, the Tsallis distribution is deemed as a superposition of the Boltzmann-Gibbs distribution. The reciprocal of the Tsallis temperature $T$ represents the average value of $1/T_{f}$. $q$ is the non-extensive parameter which measures the degree of off-equilibrium. Its deviation from unity gives the fluctuation of $1/T_{f}$\cite{q_meaning}. $\rho=\textrm{tanh}^{-1}(\beta_{S}(r/R)^{n})$ is transverse expansion rapidity, which grows as the $n$th power of the emitting source's radius ($r$). $\beta_{S}$ is the velocity of the source at the edge of the fireball ($r=R$). The mean transverse expansion velocity is $\langle \beta \rangle =2/(n+2)\beta_{S}$. In the TBW model, the default value of $n$ is  1 and $\langle \beta \rangle =2/3\beta_{S}$. In this paper, in order to investigate the dependence of the results on the choice of $n$, besides this linear velocity profile, we also consider a constant profile in which $n$ is set to be zero and $\langle \beta \rangle =\beta_{S}$. As mentioned in the introduction, this profile was applied in the generalized BGBW model when describing the identified particle's elliptic flow in Au-Au collisions\cite{TBW_v2}. For both transverse velocity profiles there are four free parameters in this model: the normalization constant, $\langle \beta \rangle$, $q$ and $T$. 

\section{Results and discussions}\label{sec:results_and_discussions}

The ALICE collaboration had presented the $\pi$, $K$, $p$, $K^{*0}$, $\phi$, $K_{S}^{0}$, $\Lambda$, $\Xi$ and $\Omega$  spectra at 0-10$\%$\footnote{For Pb-Pb and p-Pb collisions, the spectrum for a given particle at the 0-10$\%$ centrality is merged from the spectra at 0-5$\%$ and 5-10$\%$ centralities.}, 10-20$\%$\footnote{For p-Pb collisions, at 0-10$\%$ and 10-20$\%$ centralities, the spectra of $K^{*0}$ are not available so far.}, 20-40$\%$\footnote{For Pb-Pb collisions, the spectrum of $\phi$ at the 20-40$\%$(40-60$\%$) centrality is merged from the spectra at  20-30$\%$ and 30-40$\%$(40-50$\%$ and 50-60$\%$) centralities.}, 40-60$\%$ and 60-80$\%$\footnote{For Pb-Pb collisions, the spectra of $K^{*0}$ and $\phi$ at 40-60$\%$ and 60-80$\%$ centralities are taken from ref.\cite{spectra_pb_pb_2_76_kstar_phi_2}, while their spectra at other centralities are taken from ref.\cite{spectra_pb_pb_2_76_kstar_phi_1}. } centralities in Pb-Pb (p-Pb) collisions at 2.76 (5.02) TeV in refs.\cite{spectra_pb_pb_2_76_pi_k_p,spectra_pb_pb_2_76_kstar_phi_1,spectra_pb_pb_2_76_kstar_phi_2,spectra_pb_pb_2_76_kshort_lambda,spectra_pb_pb_2_76_xi_omega, LHC_3, sepectra_p_pb_kstar_phi,spectra_p_pb_xi_omega}. Here, $\pi$, $K$, $p$, $K^{*0}$, $\Xi$ and $\Omega$, respectively, refer to $\pi^{+}+\pi^{-}$, $K^{+}+K^{-}$, $p+\bar{p}$, $K^{*0}+\bar{K}^{*0}$, $\Xi^{-}+\bar{\Xi}^{+}$ and $\Omega^{-}+\bar{\Omega}^{+}$. Recently, they also published the $\pi$, $K$, $p$, $K^{*0}$ and $\phi$ spectra in Pb-Pb collisions at 5.02 TeV\cite{LHC_2, spectra_pb_pb_5_02_kstar_phi} and the $\pi$, $K$, $p$ and $\phi$ spectra in Xe-Xe collisions at 5.44 TeV\cite{spectra_Xe_Xe_5_44}.
%The spectra of $d$ and $^{3}\rm He$ in Pb-Pb (p-Pb) collisions at 2.76 (5.02) TeV are also available in refs.\cite{light_nuclei_1, spectra_p_pb_light_nuclei_1, spectra_p_pb_light_nuclei_2}.

\begin{figure}[H] 
	\centering
 \includegraphics[scale=0.25]{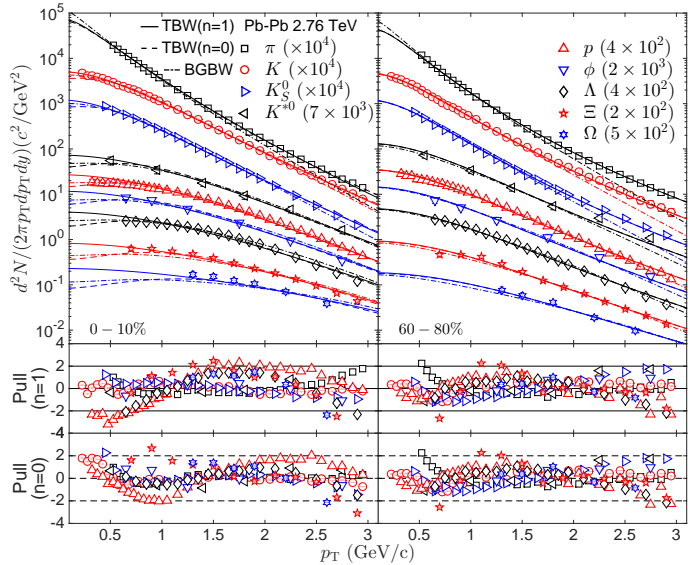}
	\caption{\label{fig:Pb_Pb_2_76_TeV_0_10_and_60_80_spectra}(Colour online) Top left (right) panel: the identified particles $p_{\rm T}$ spectra at the 0-10$\%$ (60-80$\%$) centrality in Pb-Pb collisions at 2.76 TeV. Data points are taken from refs.\cite{spectra_pb_pb_2_76_pi_k_p,spectra_pb_pb_2_76_kstar_phi_1,spectra_pb_pb_2_76_kstar_phi_2,spectra_pb_pb_2_76_kshort_lambda,spectra_pb_pb_2_76_xi_omega}. The solid (dash) curves represent the results from the TBW model with the linear (constant) velocity profile. Middle and bottom panels: the pull distributions at the 0-10$\%$ and 60-80$\%$ centralities for these two profiles.}
\end{figure}

We first perform a combined fit on the spectra of $\pi$, $K$, $p$, $K^{*0}$, $\phi$, $K_{S}^{0}$, $\Lambda$, $\Xi$ and $\Omega$ at the 0-10$\%$ centrality in Pb-Pb collisions at 2.76 TeV with the TBW model  (see equation (\ref{eq:eTBW})) using a least $\chi^{2}$ method. In the high $p_{\rm T}$ region, the hadron production is dominated by surface emission\cite{surface_emission},  which will lead to the fail of the description on the spectra with the Tsallis blast-wave model\cite{TBW_1}. Therefore, the upper limit of the $p_{\rm T}$ range for the combined fit  is set to be 3 GeV/c. In the low $p_{\rm T}$ region, a large fraction of pions originate from resonance decays\footnote{The contribution from resonance decays for the pion yield is about 16.5$\%$, which is much larger than that for the proton yield, 6$\%$, at TeV energy scale\cite{pion_proton_resonance_contribution}.}. In order to get rid of this contribution, the lower bound of the pion spectrum is chosen as 0.5 GeV/c, which is used by experimental collaborations. In the combined fit, three parameters are common for all particles: the Tsallis temperature $T$, the average transverse expansion velocity $\langle \beta \rangle$ and  the non-extensive parameter $q$. In addition, for each particle species, there is a normalization factor for its own yield. In the fit, the square root of the quadratic sum of the statistical and systematic errors of the data is utilized. The fit parameters of the model with $n=1$ and $n=0$ are both presented in Table \ref{tab:pb_pb_2_76_TeV_fit_parameters}. The $\chi^{2}$ divided by the number of degree of freedom ($\chi^{2}/$NDF) is also shown in the table. The first error in the table is the statistical uncertainty returned from the combined fit. The second one is the systematic error, which originates from the variation of the lower fit bound (from 0.5 to 0.1 GeV/c) for the pion spectrum. The same procedure is applied to the spectra of identified particles at other centralities. The parameters $T$, $\langle \beta \rangle$ and $q$ at these centralities are also listed in Table \ref{tab:pb_pb_2_76_TeV_fit_parameters}.

\begin{table}[H]
  \caption{\label{tab:pb_pb_2_76_TeV_fit_parameters}Summary of fit parameters for the TBW fit with $n=1$ and $n=0$ in Pb-Pb collisions at 2.76 TeV. See text for the explanation of the quoted uncertainties.}
  %\small
  \footnotesize
\begin{center}
  \begin{tabular}{cccccc}
\hline
\textrm{\ }&
\textrm{\ }&
\textrm{$\langle \beta \rangle$}&
\textrm{$T$ (GeV)}&
\textrm{$q$}&
\textrm{$\chi^{2}$/NDF}\\
\hline
&\textrm{0-10$\%$}& 0.565$\pm$0.005$\pm$0.009&0.096$\pm$0.003$\pm$0.012&1.032$\pm$0.007$\pm$0.019&221.295/158 \\
&\textrm{10-20$\%$}&0.550$\pm$0.005$\pm$0.012&0.097$\pm$0.003$\pm$0.014&1.040$\pm$0.006$\pm$0.022 &200.411/158\\
$n=1$&\textrm{20-40$\%$}&0.514$\pm$0.005$\pm$0.014&0.102$\pm$0.003$\pm$0.017&1.053$\pm$0.004$\pm$0.022&186.623/158\\ 
&\textrm{40-60$\%$}&0.439$\pm$0.006$\pm$0.018&0.104$\pm$0.002$\pm$0.022&1.072$\pm$0.003$\pm$0.023&117.438/158\\  
&\textrm{60-80$\%$}&0.293$\pm$0.012$\pm$0.014&0.107$\pm$0.003$\pm$0.027& 1.094$\pm$0.003$\pm$0.021&121.409/157\\   
\hline
&\textrm{0-10$\%$}& 0.614$\pm$0.004$\pm$0.008&0.087$\pm$0.002$\pm$0.011&1.073$\pm$0.002$\pm$0.012&137.022/158 \\
&\textrm{10-20$\%$}&0.594$\pm$0.004$\pm$0.009&0.089$\pm$0.002$\pm$0.013&1.076$\pm$0.002$\pm$0.013 &136.178/158\\
$n=0$&\textrm{20-40$\%$}&0.551$\pm$0.005$\pm$0.011&0.095$\pm$0.002$\pm$0.015&1.078$\pm$0.002$\pm$0.015&141.213/158\\ 
&\textrm{40-60$\%$}&0.465$\pm$0.006$\pm$0.014&0.100$\pm$0.002$\pm$0.020&1.085$\pm$0.002$\pm$0.019&97.365/158\\  
&\textrm{60-80$\%$}&0.311$\pm$0.012$\pm$0.012&0.106$\pm$0.003$\pm$0.026& 1.097$\pm$0.002$\pm$0.020&118.293/157\\   
\hline
\end{tabular}
\end{center}
\end{table}

The upper panels in Fig. \ref{fig:Pb_Pb_2_76_TeV_0_10_and_60_80_spectra} show the identified particle spectra and the combined fit results at two given centralities (0-10$\%$ and 60-80$\%$). The solid (dash) fitted curves from the TBW model with $n=1$ ($n=0$) generally describe the data well up to 3 GeV/c. In order to investigate the agreement between the model and the data points statistically, a variable $\rm pull=\rm (data-fit)/\Delta data$\footnote{$\Delta \rm data$ refers to the square root of the quadratic sum of the data's statistical and systematic errors.} is defined. The pull distributions in the middle (for $n=1$) and lower panels (for $n=0$) of the figure show that most of data agree with the fitted curves within two standard deviations.  In the region with $p_{\rm T}>$ 3 GeV/c (not shown in the figure), for the spectra of $\pi$, $K$, $K_{S}^{0}$, $\Lambda$ and $\Omega$ ($\pi$, $p$, $\Lambda$ and $\Omega$), a large descrepancy between the data and the model result with $n=1$ ($n=0$) is observed.  As a comparison, the BGBW calculations (dash-dotted curves) with $\langle \beta \rangle$ and  $T$ taken from ref.\cite{LHC_1} are also presented in the figure. It is found that the BGBW model fails to describe the spectra of $\pi$, $K$ and $K_{S}^{0}$ in the range with $p_{\rm T}>$ 1, 2 and 2 (1, 1.5 and 1.5) GeV/c at central (peripheral) collisions, respectively.

\begin{figure}[H]
	\centering
	 \includegraphics[scale=0.25]{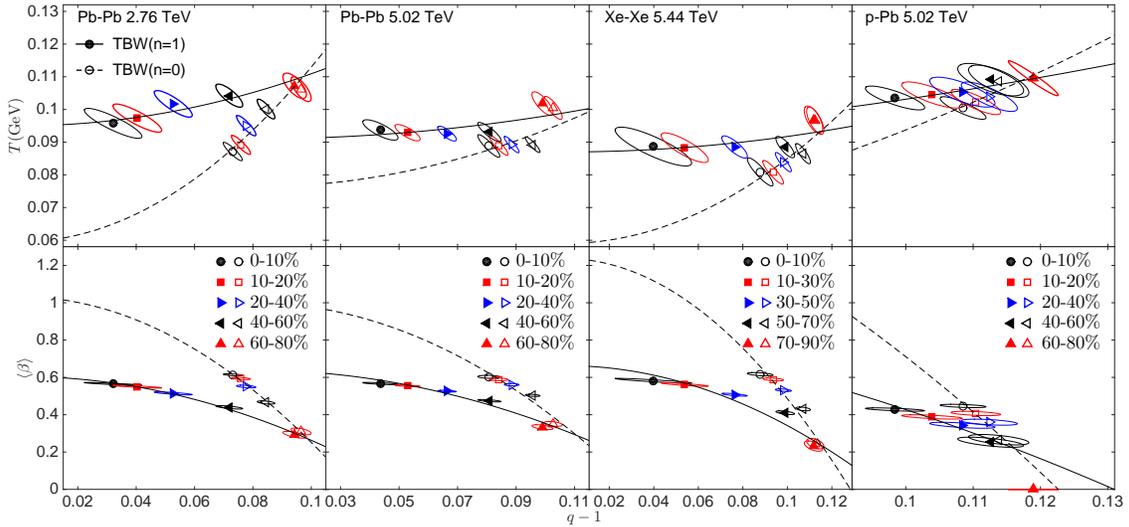}
	 \caption{\label{fig:beta_T_versus_q} (Colour online) Leftmost (middle-left, middle-right, rightmost) panel: the Tsallis temperature and the average flow velocity versus $q-1$ for Pb-Pb (Pb-Pb, Xe-Xe, p-Pb) collisions at 2.76 (5.02, 5.44, 5.02) TeV. The solid (dash) curves represent a quadratic parameterization of the parameters for the TBW model with $n=1$ ($n=0$).}
\end{figure}

Using the parameters in Table \ref{tab:pb_pb_2_76_TeV_fit_parameters},  we present $T$ and $\langle \beta \rangle$  versus $q-1$ for  Pb-Pb collisions at 2.76 TeV in the leftmost panels of Fig. \ref{fig:beta_T_versus_q}. The solid (dash) curves represent a quadratic parameterization of the parameters from the TBW fit with $n=1$ ($n=0$). The ellipse in the figure reflects the 1$\sigma$ uncertainty returned from the error matrix of the fit at a given centrality. Some conclusions can be drawn as follows.

(i)  The non-extensive parameter $q$  decreases with centrality\footnote{The dependence of $q$, $\langle \beta \rangle$ and $T$ on centrality in Pb-Pb (Xe-Xe, p-Pb) collisions at 5.02 (5.44, 5.02) is similar to that in Pb-Pb collisions at 2.76 TeV.}, which indicates that the system is more off-equilibrium in peripheral collisions than in central collisions. This trend is similar to the TBW result in Au-Au collisions at 200 GeV\cite{TBW_1}.

(ii) The average flow velocity $\langle \beta \rangle$ increases with centrality while the Tsallis temperature $T$ shows the opposite trend. This behavior is identical to the BGBW result in Pb-Pb collisions at 2.76 TeV\cite{LHC_1}. It indicates that particles are seen to decouple earlier, thus at higher temperature and with less transverse flow in peripheral collisions than in central collisions. A possible explanation is that the collision fireball in peripheral collisions does not live as long as that in central collisions and has less time to build up radial flow\cite{BW_explantion}.

(iii) For central collisions, $\langle \beta \rangle$ and $q$ ($T$) from the TBW fit with $n=1$ are smaller (is slightly larger) than those (that) from the fit with $n=0$. For peripheral collisions, $\langle \beta \rangle$, $q$ and $T$ from the former are compatible with the corresponding values from the latter\footnote{For Pb-Pb (Xe-Xe, p-Pb) collisions at 5.02 (5.44, 5.02) TeV, similar conclusions are obtained from the comparison of parameters returned from the TBW fit with $n=1$ and $n=0$. }.

(iv) The dependence of $\langle \beta \rangle$ and $T$ on $q-1$ is  nonlinear and has a negative correlation. For the model with $n=1$ ($n=0$), it is parameterized with a quadratic distribution: $\langle \beta \rangle=(0.605\pm0.005)-(34.215\pm1.349)(q-1)^{2}$ and $T=(0.095\pm0.001)+(1.592\pm0.034)(q-1)^{2}$ ($\langle \beta \rangle=(1.033\pm0.026)-(78.234\pm4.185)(q-1)^{2}$ and $T=(0.059\pm0.007)+(5.322\pm1.138)(q-1)^{2}$). For  both $\langle \beta \rangle$ and $T$, this dependence for $n=1$ is weaker than that for $n=0$\footnote{For Pb-Pb (Xe-Xe, p-Pb) collisions at 5.02 (5.44, 5.02) TeV, similar results are obtained from the comparison between the nonlinear dependence of parameters in the TBW model with $n=1$ and $n=0$. }. 

\setcounter{footnote}{0}

\begin{figure}[H]
 	\centering
 	 \includegraphics[scale=0.25]{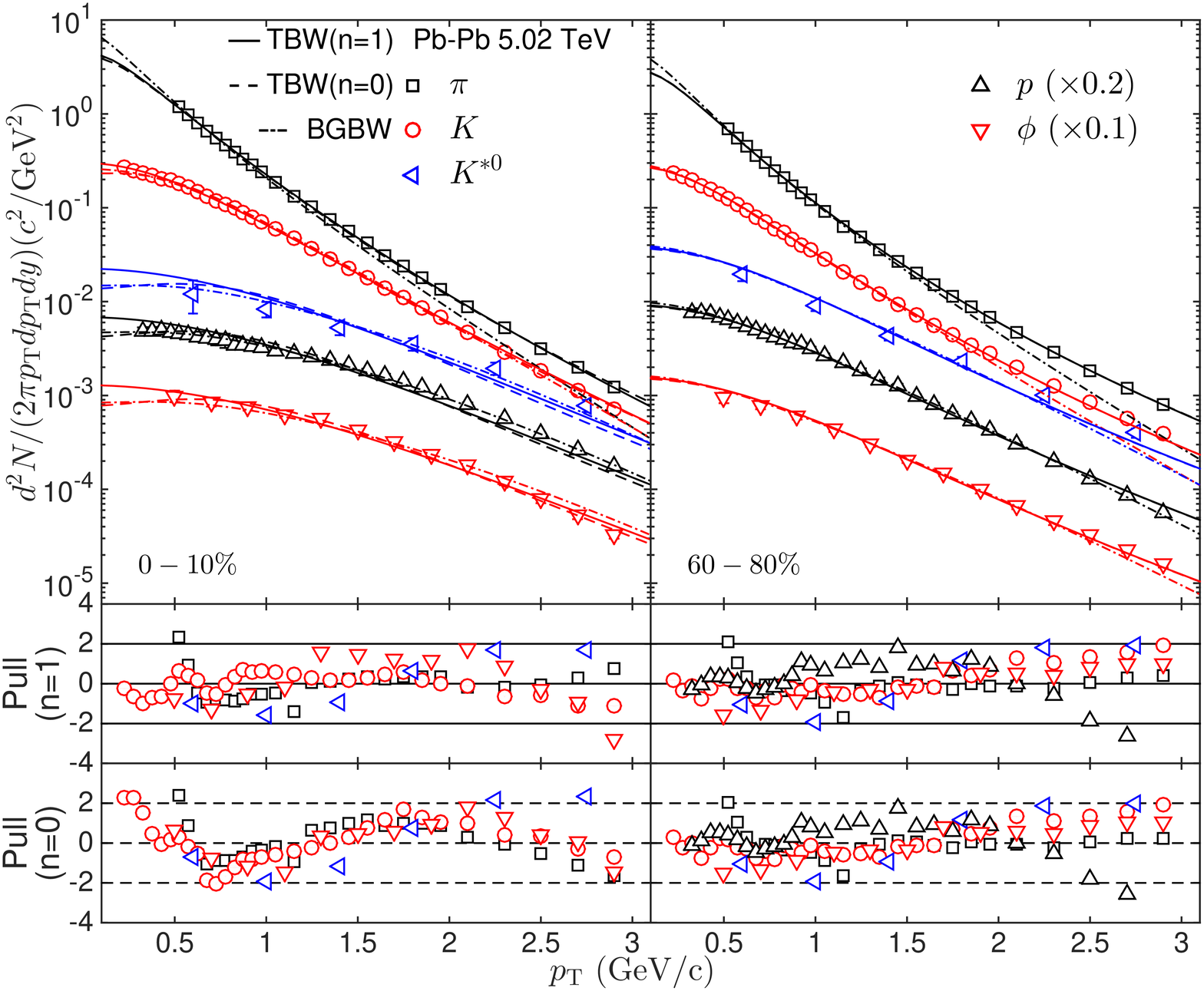}
 	 \caption{\label{fig:Pb_Pb_5_02_TeV_0_10_and_60_80_spectra}(Colour online) Top left (right) panel: the identified particles $p_{\rm T}$ spectra at the 0-10$\%$ (60-80$\%$) centrality in Pb-Pb collisions at 5.02 TeV. Data points are taken from refs.\cite{LHC_2, spectra_pb_pb_5_02_kstar_phi}. The solid (dash) curves represent the results from the TBW model with the linear (constant) velocity profile. Middle and bottom panels: the pull distributions at the 0-10$\%$ and 60-80$\%$ centralities for these two profiles.}
\end{figure}

Next we investigate the spectra of $\pi$, $K$, $p$, $K^{*0}$ and $\phi$ in Pb-Pb collisions at 5.02 TeV with the TBW model.  Very poor $\chi^{2}/$NDF, 3.37, 2.93 and 2.12 (3.79, 2.82 and 1.81) are obtained when the TBW model with $n=1$ ($n=0$) is, respectively, applied to these spectra at the 0-10$\%$\footnote{The spectra of $\pi$, $K$ and $p$ at the 0-10$\%$ centrality are merged from the spectra at 0-5$\%$ and 5-10$\%$ centralities.}, 10-20$\%$ and 20-40$\%$\footnote{The spectra of $\pi$, $K$, $p$, $K^{*0}$ and $\phi$ at the 20-40$\%$ (40-60$\%$, 60-80$\%$) centrality are merged from the spectra at 20-30$\%$ and 30-40$\%$ (40-50$\%$ and 50-60$\%$, 60-70$\%$ and 70-80$\%$) centralities.} centralities. However, with the exclusion of the proton spectrum, the performance of the combined fit at these centralities becomes very good. The upper panels of Fig. \ref{fig:Pb_Pb_5_02_TeV_0_10_and_60_80_spectra} show the spectra of identified particles with their associated TBW results at two selected centralities (0-10$\%$ and 60-80$\%$). The data are generally depicted by the model (the solid curves for $n=1$ and the dash curves for $n=0$) well.  Judging from the pull distributions in the middle and lower panels of the figure, at both the 0-10$\%$ and 60-80$\%$ centralities most of the data are in agreement with the model results within 2 standard deviations. Also presented in the figure are the BGBW calculations (dash-dotted curves) with $\langle \beta \rangle$ and  $T$ taken from ref.\cite{LHC_2}. For $\pi$ and $K$ in central (peripheral) collisions, the BGBW model fails to describe their spectra in the region with $p_{\rm T}>$ 1 and 2 (1 and 1.5) GeV/c, respectively. For $K^{*0}$ and protons in the 0-10$\%$ centrality, the BGBW curve well matches the data up to 3 GeV/c, while the TBW model with both $n=1$ and $n=0$ underestimates the spectra with $p_{\rm T}$ above 1.8 and 1.55 GeV/c. This possibly implies that $K^{*0}$ and protons prefer to freeze out earlier and thus less off-equilibrium than other light hadrons in central collisions. The fit parameters, their uncertainties and the $\chi^{2}/$NDF are reported in Table \ref{tab:pb_pb_5_02_TeV_fit_parameters}. For the TBW model with $n=1$ or $n=0$, when considering the uncertainties, in central collisions $\langle \beta \rangle$ and $T$ are comparable with the corresponding values in Pb-Pb collisions at 2.76 TeV\footnote{In both central and peripheral Pb-Pb collisions, the degree of non-equilibrium at 5.02 TeV is compatible with that at 2.76 TeV when considering the parameter's uncertainty. }.  However, in peripheral collisions, $\langle \beta \rangle$ ($T$) is larger (slightly smaller) than that in Pb-Pb collisions at 2.76 TeV.

\begin{table}[H]
  \caption{\label{tab:pb_pb_5_02_TeV_fit_parameters}Summary of fit parameters for the TBW fit with $n=1$ and $n=0$ in Pb-Pb collisions at 5.02 TeV.  The explanation for the uncertainties is the same as that in Table \ref{tab:pb_pb_2_76_TeV_fit_parameters}.}
\small
\begin{center}
  \begin{tabular}{cccccc}
\hline
\textrm{\ }&
\textrm{\ }&
\textrm{$\langle \beta \rangle$}&
\textrm{$T$ (GeV)}&
\textrm{$q$}&
\textrm{$\chi^{2}$/NDF}\\
\hline
&\textrm{0-10$\%$}& 0.567$\pm$0.004$\pm$0.010&0.094$\pm$0.002$\pm$0.019&1.044$\pm$0.004$\pm$0.029&57.12/68 \\
&\textrm{10-20$\%$}&0.555$\pm$0.003$\pm$0.009&0.093$\pm$0.002$\pm$0.019&1.053$\pm$0.003$\pm$0.027 &38.90/68\\
$n=1$&\textrm{20-40$\%$}&0.526$\pm$0.003$\pm$0.007&0.093$\pm$0.002$\pm$0.019&1.066$\pm$0.002$\pm$0.024&33.28/68\\ 
&\textrm{40-60$\%$}&0.475$\pm$0.004$\pm$0.017&0.093$\pm$0.002$\pm$0.019&1.081$\pm$0.003$\pm$0.023&117.16/96\\  
&\textrm{60-80$\%$}&0.333$\pm$0.009$\pm$0.013&0.102$\pm$0.002$\pm$0.029& 1.099$\pm$0.002$\pm$0.025&93.39/96\\   
\hline
&\textrm{0-10$\%$}& 0.603$\pm$0.005$\pm$0.006&0.089$\pm$0.002$\pm$0.018&1.081$\pm$0.002$\pm$0.020&89.72/68 \\
&\textrm{10-20$\%$}&0.589$\pm$0.005$\pm$0.005&0.089$\pm$0.002$\pm$0.018&1.084$\pm$0.002$\pm$0.020 &69.19/68\\
$n=0$&\textrm{20-40$\%$}&0.561$\pm$0.004$\pm$0.005&0.089$\pm$0.002$\pm$0.018&1.089$\pm$0.002$\pm$0.019&46.41/68\\ 
&\textrm{40-60$\%$}&0.504$\pm$0.004$\pm$0.013&0.089$\pm$0.002$\pm$0.017&1.096$\pm$0.002$\pm$0.018&89.26/96\\  
&\textrm{60-80$\%$}&0.351$\pm$0.009$\pm$0.009&0.100$\pm$0.002$\pm$0.028& 1.103$\pm$0.002$\pm$0.023&90.69/96\\   
\hline
\end{tabular}
\end{center}
\end{table}

The nonlinear dependence of $\langle \beta \rangle$ and $T$ on $q-1$ in Pb-Pb collisions at 5.02 TeV is shown in the middle-left panels of Fig. \ref{fig:beta_T_versus_q}.  For the parameters from the TBW fit with $n=1$ ($n=0$), the dependence is parameterized as $\langle \beta \rangle=(0.639\pm0.019)-(28.562\pm3.920)(q-1)^{2}$ and $T=(0.091\pm$ $0.002)+(0.702\pm0.465)(q-1)^{2}$ ($\langle \beta \rangle=(1.000\pm0.076)-(58.040\pm9.856)(q-1)^{2}$ and $T=(0.076\pm$ $0.009)+(1.744\pm1.092)(q-1)^{2}$). For both $\langle \beta \rangle$ and $T$,  this dependence is weaker than that in Pb-Pb collisions at 2.76 TeV.

\begin{figure}[H]
  	\centering
	 \includegraphics[scale=0.25]{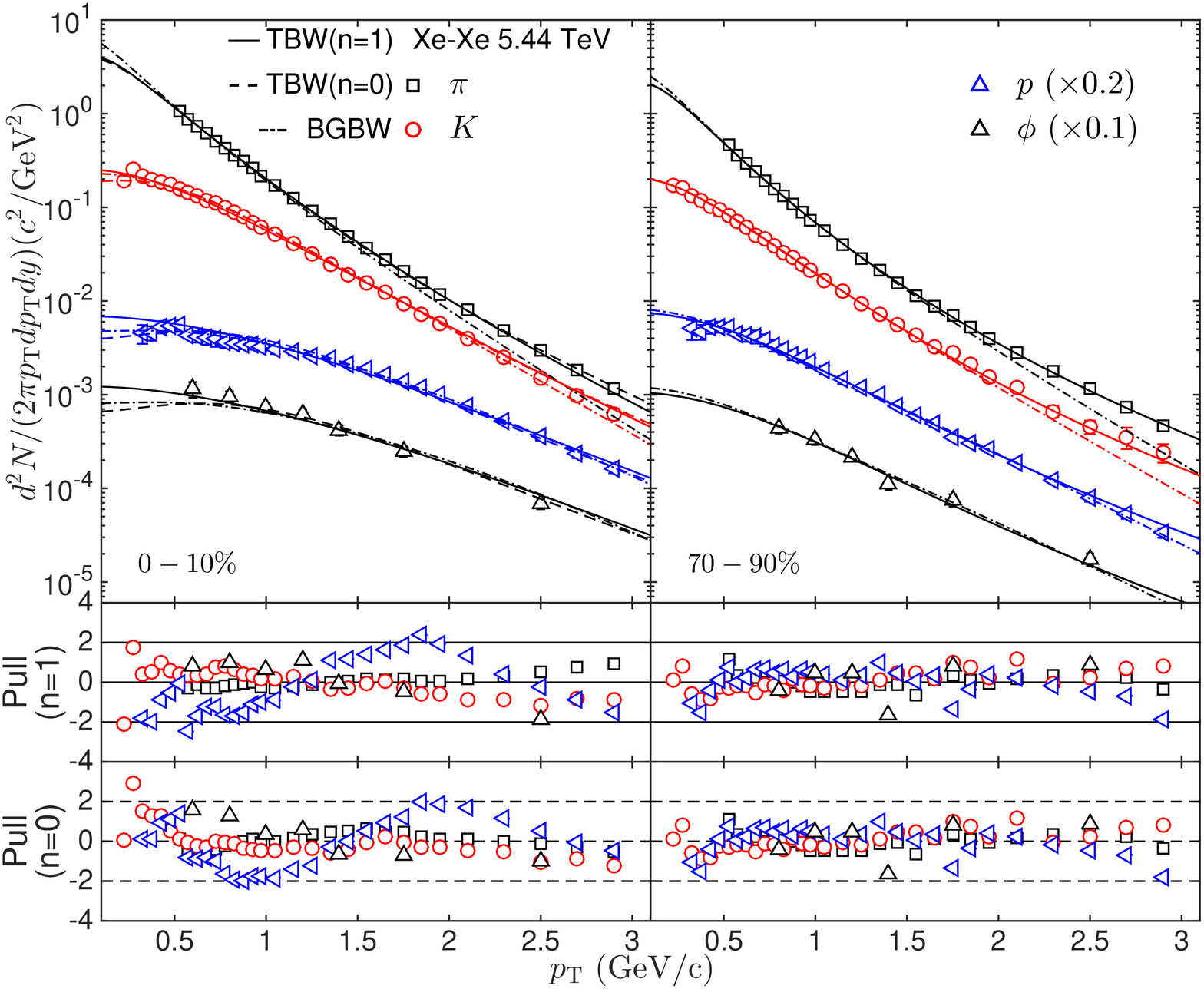}
  	 \caption{\label{fig:Xe_Xe_5_44_TeV_0_10_and_70_90_spectra}(Colour online) Top left (right) panel: the  identified particles $p_{\rm T}$ spectra at the 0-10$\%$ (70-90$\%$) centrality in Xe-Xe collisions at 5.44 TeV. Data points are taken from ref.\cite{spectra_Xe_Xe_5_44}. The solid (dash) curves represent the results from the TBW model with the linear (constant) velocity profile. Middle and bottom panels: the pull distributions at the 0-10$\%$ and 70-90$\%$ centralities for these two profiles.}
\end{figure}

In order to compare the hydrodynamic  expansion in systems at similar charged particle multiplicities but with different initial geometrical eccentricities, we extend the study to the spectra of $\pi$, $K$, $p$ and $\phi$ in Xe-Xe collisions at 5.44 TeV\footnote{The spectra of $\pi$, $K$ and $p$ at the 0-10$\%$ (10-30$\%$, 30-50$\%$, 50-70$\%$) centrality are merged from the spectra at 0-5$\%$ and 5-10$\%$ (10-20$\%$ and 20-30$\%$, 30-40$\%$ and 40-50$\%$, 50-60$\%$ and 60-70$\%$) centralities.}.  The upper panels of Fig. \ref{fig:Xe_Xe_5_44_TeV_0_10_and_70_90_spectra} present the spectra of identified particles with their associated TBW results at two given centralities (0-10$\%$ and 70-90$\%$). The TBW model with $n=1$ (solid curves) or $n=0$ (dash curves) generally reproduces the spectra.  The pull distributions in the middle and lower panels of the figure show that most of the data are consistent with the model results within 2 standard deviations. Also displayed in the figure are the BGBW calculations (dash-dotted curves)\footnote{For Xe-Xe collisions, the values of $\langle \beta \rangle$ and $T$ in the BGBW model are not available so far in literature. Thus we performed a combined fit to the $\pi$, $K$ and $p$ spectra in the ranges 0.5–1 GeV/c, 0.2–1.5 GeV/c and 0.3–3 GeV/c, respectively. They are 0.657$\pm$0.004 and 0.097$\pm$0.004 GeV (0.427$\pm$0.010 and 0.143$\pm$0.005 GeV) for central (peripheral) collisions. }. For $\pi$ and $K$ in central (peripheral) collisions, the BGBW model underestimates their spectra in the region with $p_{\rm T}>$ 1 and 2 (1 and 1.5) GeV/c, respectively.  The fit parameters, their uncertainties and the $\chi^{2}/$NDF are tabulated in Table \ref{tab:Xe_Xe_5_44_TeV_fit_parameters}. In ref.\cite{multiplicity_Xe_Xe_5_44}, it showed that the central (0-10$\%$) Xe-Xe collisions have similar multiplicity as the semi-central (10-20$\%$) Pb-Pb collisions at 5.02 TeV. However, the average flow velocity returned from the TBW model in the former case is slightly larger than that in the latter case. As described in that reference, the radial flow of the system is mainly driven by the multiplicity and not by the collision geometry while the elliptic flow is dominantly influenced by the initial eccentricity. Thus, we infer that the discrepancy between the flow velocities in Xe-Xe and Pb-Pb collisions at similar multiplicities is due to the difference of the colliding energy. 

The nonlinear dependence of $\langle \beta \rangle$ and $T$ on $q-1$ in Xe-Xe collisions at 5.44 TeV is presented in the middle-right panels of Fig. \ref{fig:beta_T_versus_q}.  For the TBW model with $n=1$ ($n=0$), the parameterization of the dependence is $\langle \beta \rangle=(0.663\pm0.019)-(32.137\pm4.913)(q-1)^{2}$ and $T=(0.087\pm$ $0.002)+(0.473\pm0.315)(q-1)^{2}$ ($\langle \beta \rangle=(1.237\pm0.123)-(75.119\pm11.761)(q-1)^{2}$ and $T=(0.059\pm$ $0.007)+(2.600\pm0.724)(q-1)^{2}$). For both $\langle \beta \rangle$ and $T$,  this dependence is compatible with that in Pb-Pb collisions at 5.02 TeV when considering the uncertainties of the coefficient for $(q-1)^2$.

\begin{table}[H]
  \caption{\label{tab:Xe_Xe_5_44_TeV_fit_parameters}Summary of parameters for the TBW fit with $n=1$ and $n=0$  in Xe-Xe collisions at 5.44 TeV.  The explanation for the uncertainties is the same as that in Table \ref{tab:pb_pb_2_76_TeV_fit_parameters}.}
\small
\begin{center}
  \begin{tabular}{cccccc}
\hline
\textrm{\ }&
\textrm{\ }&
\textrm{$\langle \beta \rangle$}&
\textrm{$T$ (GeV)}&
\textrm{$q$}&
\textrm{$\chi^{2}$/NDF}\\
\hline
&\textrm{0-10$\%$}& 0.582$\pm$0.009$\pm$0.027&0.089$\pm$0.004$\pm$0.024&1.040$\pm$0.011$\pm$0.049&83.56/85 \\
&\textrm{10-30$\%$}&0.562$\pm$0.006$\pm$0.030&0.088$\pm$0.003$\pm$0.025&1.054$\pm$0.007$\pm$0.047 &53.14/85\\
$n=1$&\textrm{30-50$\%$}&0.507$\pm$0.005$\pm$0.029&0.089$\pm$0.002$\pm$0.026&1.076$\pm$0.004$\pm$0.039&31.16/85\\ 
&\textrm{50-70$\%$}&0.411$\pm$0.007$\pm$0.026&0.089$\pm$0.002$\pm$0.026&1.099$\pm$0.003$\pm$0.031&26.15/85\\  
&\textrm{70-90$\%$}&0.232$\pm$0.018$\pm$0.006&0.097$\pm$0.003$\pm$0.033& 1.112$\pm$0.003$\pm$0.026&30.63/84\\   
\hline
&\textrm{0-10$\%$}& 0.617$\pm$0.006$\pm$0.025&0.081$\pm$0.003$\pm$0.020&1.088$\pm$0.004$\pm$0.028&75.29/85\\
&\textrm{10-30$\%$}&0.589$\pm$0.005$\pm$0.023&0.081$\pm$0.002$\pm$0.021&1.094$\pm$0.003$\pm$0.027 &48.80/85\\
$n=0$&\textrm{30-50$\%$}&0.531$\pm$0.005$\pm$0.023&0.084$\pm$0.002$\pm$0.023&1.098$\pm$0.002$\pm$0.028&27.49/85\\ 
&\textrm{50-70$\%$}&0.431$\pm$0.006$\pm$0.022&0.087$\pm$0.002$\pm$0.025&1.107$\pm$0.002$\pm$0.027&24.72/85\\  
&\textrm{70-90$\%$}&0.245$\pm$0.018$\pm$0.009&0.097$\pm$0.003$\pm$0.033& 1.113$\pm$0.003$\pm$0.026&30.47/84\\   
\hline
\end{tabular}
\end{center}
\end{table}

\begin{figure}[H]
  	\centering
	 \includegraphics[scale=0.25]{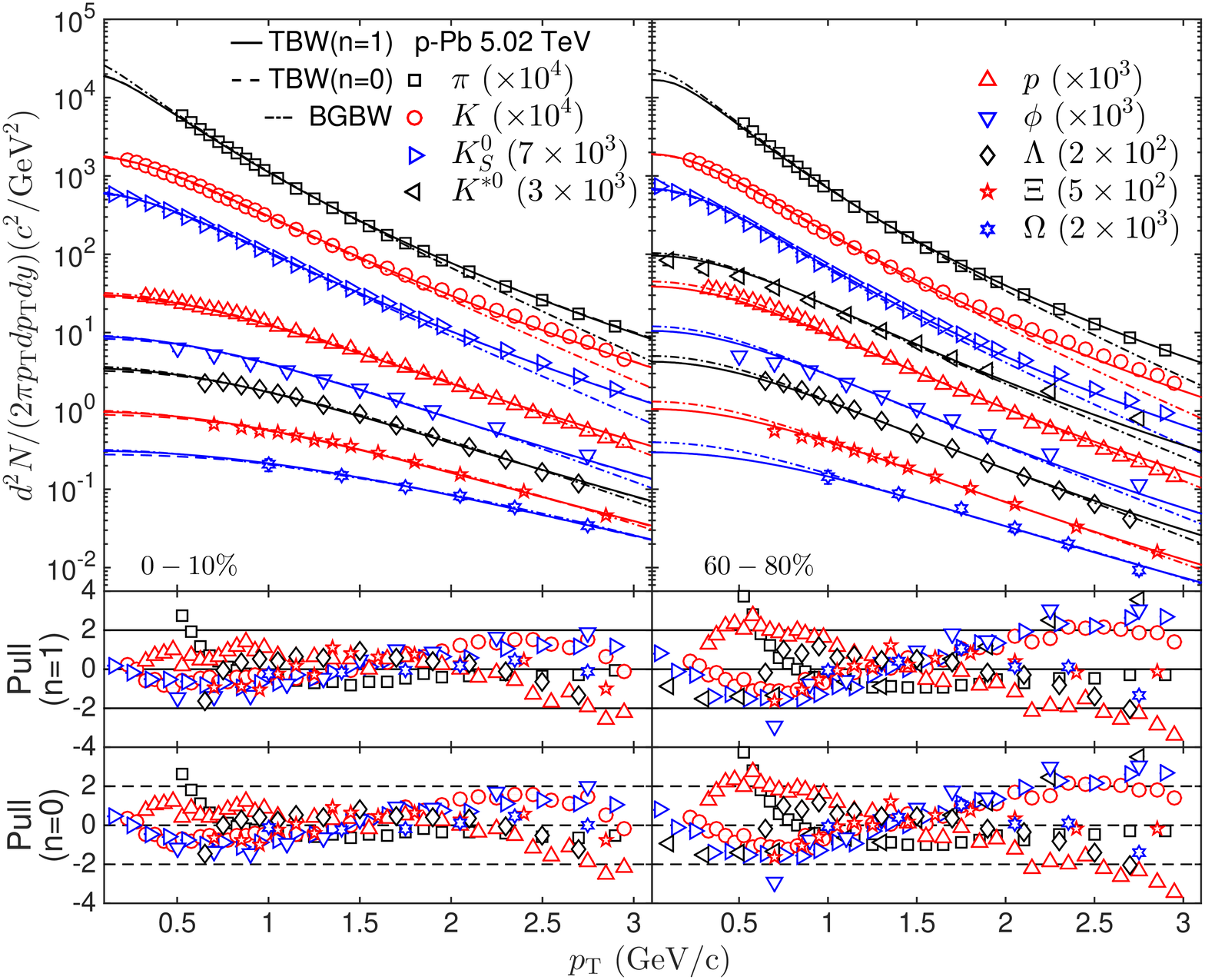}
  	 \caption{\label{fig:p_Pb_5_02_TeV_0_10_and_60_80_spectra}(Colour online) Top left (right) panel: the identified particles $p_{\rm T}$ spectra at the 0-10$\%$ (60-80$\%$) centrality in p-Pb collisions at 5.02 TeV. Data points are taken from refs.\cite{LHC_3, sepectra_p_pb_kstar_phi,spectra_p_pb_xi_omega}. The solid (dash) curves represent the results from the TBW model with the linear (constant) velocity profile. Middle and bottom panels: the pull distributions at the 0-10$\%$ and 60-80$\%$ centralities for these two profiles.}
\end{figure}

As the size of the system in pA collisions is between those in pp and AA collisions, the data in pA collisions has usually been used as a reference to separate initial state (cold nuclear matter) effects from final state (hot and dense matter) effects\cite{LHC_3}. Thus, we extend the investigation to the spectra of $\pi$, $K$, $p$, $K^{*0}$, $\phi$, $K_{S}^{0}$, $\Lambda$, $\Xi$ and $\Omega$ at a given centrality in p-Pb collisions at 5.02 TeV. In the upper panels of Fig. \ref{fig:p_Pb_5_02_TeV_0_10_and_60_80_spectra}, the identified particles spectra at two selected centralities (0-10$\%$ and 60-80$\%$) are presented together with the results (solid and dash curves) from the TBW model with $n=1$ and $n=0$. From the pull distributions in the middle and lower panels of the figure, we find that most of the data agree with the model within 2 standard deviations. Also shown in the figure are the BGBW calculations (dash-dotted curves) with $\langle \beta \rangle$ and  $T$ taken from ref.\cite{LHC_3}. For $\pi$, $K$ and $K_{S}^{0}$, the BGBW model can not reproduce their spectra in the range with $p_{\rm T}>$ 1.5 GeV/c. The parameters for the TBW fit with both $n=1$ and $n=0$ are presented in Table \ref{tab:p_pb_5_02_TeV_fit_parameters}. For the particle spectra at the 60-80$\%$ centrality, $\langle \beta \rangle$ returned from the fit with $n=1$ ($n=0$) is $2.7 \times 10^{-5}$ ($5.5 \times 10^{-6}$). Thus we set it to be 0 and repeat the analysis procedure. For the TBW model with $n=1$($n=0$),  in central collisions $\langle \beta \rangle$  is  significantly less than that in Pb-Pb collisions at 5.02 TeV. Moreover, it is larger than that  in proton-proton (pp) collisions at 5.02 TeV\cite{pp_5_02_pi_k_p}, 0.224$\pm0.012$ (0.231$\pm0.013$)\footnote{This value is returned from the combined fit of the TBW model with $n=1$ ($n=0$) to the $\pi$, $K$ and $p$ spectra in pp collisions at 5.02 TeV.}.  This could be interpreted that the radial flow in p-Pb collisions is mainly due to the cold nuclear matter effect. In Pb-Pb collisions, not only the cold nuclear matter effect  but also the hot dense matter effect exists, which will make the radial flow more stronger.

 The nonlinear dependence of $\langle \beta \rangle$ and $T$ on $q-1$ in p-Pb collisions at 5.02 TeV is exhibited in the rightmost panels of Fig. \ref{fig:beta_T_versus_q}. For the TBW model with $n=1$ ($n=0$), this dependence is fitted with the distribution $\langle \beta \rangle=(1.025\pm0.130)-(59.773\pm11.172)(q-1)^{2}$ and $T=(0.088\pm0.004)+(1.514\pm0.303)(q-1)^{2}$ ($\langle \beta \rangle=(2.121\pm0.406)-(141.025\pm32.688)(q-1)^{2}$ and $T=(0.053\pm0.012)+(4.054\pm0.970)(q-1)^{2}$). This dependence is stronger than that in Pb-Pb collisions at 5.02 TeV.

 \begin{table}[H]
   \caption{\label{tab:p_pb_5_02_TeV_fit_parameters} Summary of parameters for the TBW fit with $n=1$ and $n=0$  in p-Pb collisions at 5.02 TeV.  The explanation for the uncertainties is the same as that in Table \ref{tab:pb_pb_2_76_TeV_fit_parameters}.}
%\small
\footnotesize
\begin{center}
  \begin{tabular}{cccccc}
\hline
\textrm{\ }&
\textrm{\ }&
\textrm{$\langle \beta \rangle$}&
\textrm{$T$ (GeV)}&
\textrm{$q$}&
\textrm{$\chi^{2}$/dof}\\
\hline
& \textrm{0-10$\%$}& 0.428$\pm$0.006$\pm$0.026&0.104$\pm$0.002$\pm$0.039&1.098$\pm$0.003$\pm$0.040&116.62/152 \\
& \textrm{10-20$\%$}&0.388$\pm$0.008$\pm$0.023&0.104$\pm$0.003$\pm$0.039&1.104$\pm$0.003$\pm$0.037&143.32/152\\
$n=1$& \textrm{20-40$\%$}&0.343$\pm$0.011$\pm$0.016&0.105$\pm$0.003$\pm$0.040&1.108$\pm$0.003$\pm$0.034&215.59/163\\ 
& \textrm{40-60$\%$}&0.253$\pm$0.018$\pm$0.003&0.109$\pm$0.003$\pm$0.041&1.113$\pm$0.003$\pm$0.030&260.83/163\\  
& \textrm{60-80$\%$}&0 (fixed)&0.110$\pm$0.004$\pm$0.039& 1.119$\pm$0.002$\pm$0.026&347.58/164\\   
\hline
& \textrm{0-10$\%$}& 0.447$\pm$0.006$\pm$0.019&0.100$\pm$0.002$\pm$0.037&1.109$\pm$0.002$\pm$0.034&113.39/152 \\
& \textrm{10-20$\%$}&0.407$\pm$0.008$\pm$0.017&0.102$\pm$0.003$\pm$0.038&1.110$\pm$0.002$\pm$0.033 &139.68/152\\
$n=0$& \textrm{20-40$\%$}&0.360$\pm$0.011$\pm$0.011&0.104$\pm$0.003$\pm$0.039&1.112$\pm$0.003$\pm$0.032&213.35/163\\ 
& \textrm{40-60$\%$}&0.266$\pm$0.018$\pm$0.007&0.109$\pm$0.003$\pm$0.041&1.114$\pm$0.003$\pm$0.029&260.18/163\\  
& \textrm{60-80$\%$}&0 (fixed)&0.110$\pm$0.004$\pm$0.039& 1.119$\pm$0.002$\pm$0.026&347.58/164\\   
\hline
\end{tabular}
\end{center}
 \end{table}

\begin{figure}[H]
	\centering
	\includegraphics[scale=0.3]{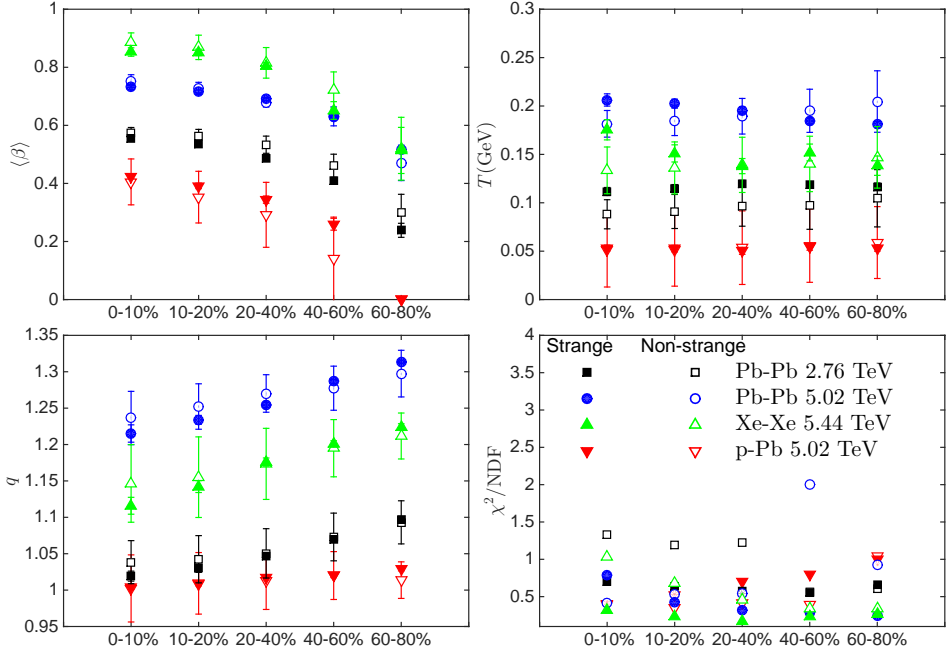}
	\caption{\label{fig:strange_non_strange_parameters} $\langle \beta \rangle$, $T$, $q$ and $\chi^{2}$/NDF from the TBW fit with $n=1$ as a function of centrality for strange and non-strange  hadrons in Pb-Pb (Pb-Pb, Xe-Xe, p-Pb) collisions at 2.76 (5.02, 5.44, 5.02) TeV. The error bar represents the parameter's total uncertainty. }
\end{figure}

As shown in ref.\cite{TBW_2}, in Au-Au collisions at 200 GeV, for central collisions the radial flow (the Tsallis temperature) for strange hadrons is smaller (larger)  than light hadrons without strange content, indicating that strange hadrons decouple from the system earlier than non-strange hadrons. It is presumed that a similar scenario happens in heavy-ion collisions at higher energies. Thus we extend our investigation to the spectra of strange and non-strange hadrons in Pb-Pb (Pb-Pb, Xe-Xe, p-Pb) collisions at 2.76 (5.02, 5.44, 5.02) TeV. Fig. \ref{fig:strange_non_strange_parameters} presents $\langle \beta \rangle$, $T$, $q$ and $\chi^{2}$/NDF in the TBW model with $n=1$ as a function of centrality for different colliding systems.  In order to increase the visibility, for both strange and non-strange hadrons, $\langle \beta \rangle$  in Pb-Pb (Xe-Xe) collisions at 5.02 (5.44) TeV is shifted upwards by 0.15 (0.3), while $q$ and $T$ in Pb-Pb (Xe-Xe, p-Pb) collisions at 5.02 (5.44, 5.02) TeV are shifted upwards (upwards, downwards) by 0.2 and 0.1 (0.1 and 0.05, 0.1 and 0.05), respectively. For Pb-Pb collisions at 2.76 TeV, in central (peripheral) collisions, $\langle \beta \rangle$, $T$ and $q$ of strange hadrons are slightly smaller, larger and smaller (smaller, larger and larger) than those of non-strange hadrons. The differences are around 1.0$\sigma$, 1.6$\sigma$ and 0.6$\sigma$ (1.0$\sigma$, 0.4$\sigma$ and 0.1$\sigma$), respectively.  For Pb-Pb collisions at 5.02 TeV, in central (peripheral) collisions, strange hadrons have slightly smaller, larger and smaller (larger, smaller and larger) $\langle \beta \rangle$, $T$ and $q$ than those of non-strange hadrons. The differences are, respectively, about 1.0$\sigma$, 1.8$\sigma$ and 0.6$\sigma$ (0.7$\sigma$, 0.7$\sigma$ and 0.5$\sigma$). For Xe-Xe collisions at 5.44 TeV, in central (peripheral) collisions, $\langle \beta \rangle$, $T$ and $q$ of strange hadrons are slightly smaller, larger and smaller (smaller, smaller and larger)  than those of non-strange hadrons. The differences are, respectively, about 1.1$\sigma$, 1.7$\sigma$ and 0.6$\sigma$ (0.05$\sigma$, 0.3$\sigma$ and 0.4$\sigma$).  For p-Pb collisions at 5.02 TeV, in central (peripheral) collisions, $\langle \beta \rangle$, $T$ and $q$ of strange hadrons are slightly larger, larger and larger than (equal to, smaller than and larger than) those of non-strange hadrons. The differences are, respectively, 0.2$\sigma$, 0.06$\sigma$ and 0.03$\sigma$ (0$\sigma$, 0.6$\sigma$ and 0.2$\sigma$). A possible explanation is as follows. For Pb-Pb (Pb-Pb, Xe-Xe) central collisions, strange hadrons prefer to freeze out earlier than non-strange hadrons, as the formers are generally heavier than the latters. However, in peripheral collisions, the freeze-out time of strange hadrons is very close to that of non-strange hadrons, because the collision system is small and only can exist in a short time. For p-Pb collisions at 5.02 TeV, as the difference of the parameters for strange and non-strange hadrons is not profound, it is difficult to make a definite conclusion about the freeze-out order for this system. Similar conclusions can be made from the comparison between the parameters of strange and non-strange hadrons from the TBW model with $n=0$.

\begin{figure}[H]
	\centering
	\includegraphics[scale=0.3]{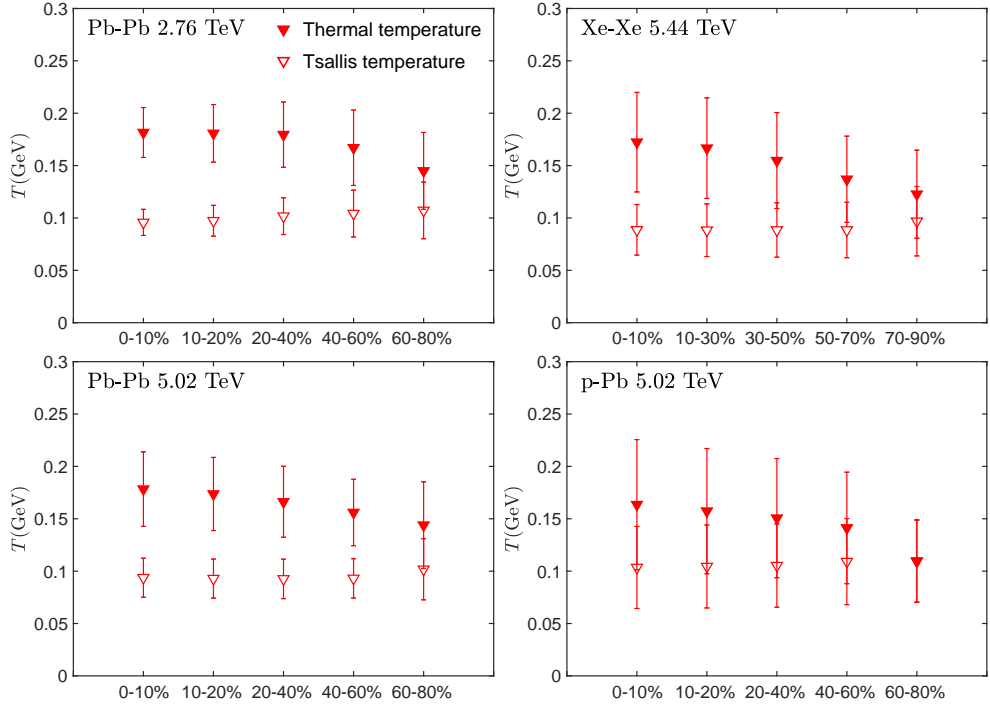}
	\caption{\label{fig:thermal_temperature} The thermal temperature and the Tsallis temperature for the TBW fit with $n=1$ as a function of centrality for identified hadrons in Pb-Pb (Pb-Pb, Xe-Xe, p-Pb) collisions at 2.76 (5.02, 5.44, 5.02) TeV. The error bar represents the parameter's total uncertainty. }
\end{figure}

Finally, as described in ref.\cite{BGBW_dist}, the temperature obtained directly from the BGBW fit, $T_{f}$, does not equal to the thermal temperature\footnote{It was defined as the inverse slope at high $m_{T}$ or $p_{T}$.}, $T_{th}$, at the light hadrons freeze-out. Usually it results in a blue shift compared to the original BGBW temperature due to the existence of a radial flow, $T_{th}=T_{f}\sqrt{(1+\langle \beta \rangle)/(1-\langle \beta \rangle)}$. Similarly, in the TBW model, the thermal temperature is also larger than the Tsallis temperature, $T$, by a blue shift factor,  $T_{th}=T\sqrt{(1+\langle \beta \rangle)/(1-\langle \beta \rangle)}$\footnote{The derivation of this formula for the TBW model can be found in the appendix.}, which is the same as in the BGBW model. In Fig. \ref{fig:thermal_temperature}, we present the thermal temperature for the TBW model with $n=1$ as a function of centrality for Pb-Pb (Pb-Pb, Xe-Xe, p-Pb) collisions at 2.76 (5.02, 5.44, 5.02) TeV. As a comparison, the Tsallis temperature is also shown in the figure. We observe that the thermal temperature decreases from central to peripheral collisions, exhibiting the same trend as the average transverse momentum\cite{LHC_1,LHC_3,spectra_Xe_Xe_5_44} while showing an opposite behavior to the Tsallis temperature. Moreover, the former is systematically higher than the latter, except for the p-Pb collisions at the 60-80$\%$ centrality where the average flow velocity is 0. The difference between these two temperatures is larger in central collisions than that in peripheral collisions. The reason is obvious: the average flow velocity is larger in more central collisions, which leads to a larger blue shift factor. 

\section{Conclusions}
\label{sec:conclusions}

In summary, a comprehensive investigation of the identified particle spectra in Pb-Pb (Pb-Pb, Xe-Xe, p-Pb) collisions at 2.76 (5.02, 5.44, 5.02) TeV is performed via the TBW model with the linear as well as the constant velocity profile. The model generally reproduces the spectra well up to 3 GeV/c. We observe that $\langle \beta \rangle$ decreases from central to peripheral collisions while the $T$ and $q$ show the opposite behavior, indicating a more rapid expansion and less off-equilibrium  of the system in more central collisions. A possible explanation is that the collision fireball in peripheral collisions does not live as long as that in central collisions and has less time to build up the radial flow and to reach the equilibrium. Moreover, we find that in central collisions $\langle \beta \rangle$ and $q$ ($T$) from the model with the linear profile are smaller (is slightly larger) than those (that) from the model with the constant profile, while in peripheral collisions $\langle \beta \rangle$, $T$ and $q$ from the former are compatible with those from the latter. For Pb-Pb (Pb-Pb, Xe-Xe) collisions at 2.76 (5.02, 5.44) TeV, in central collisions, strange hadrons tend to freeze out earlier than non-strange hadrons, while in peripheral collisions they prefer to freeze out at similar time. For p-Pb collisions at 5.02 TeV, as the difference of parameters for strange and non-strange hadrons is not profound, the picture of the freeze-out order is unclear. Finally, we see that in Pb-Pb peripheral collisions $\langle \beta \rangle$ ($T$) at 5.02 TeV prefers to be larger (smaller) than that at 2.76 TeV. Similar difference appears in the comparison of parameters at a given centrality in Pb-Pb and p-Pb collisions at 5.02 TeV, and in the comparison of parameters at centralities with similar multiplicities in Xe-Xe collisions at 5.44 TeV and Pb-Pb collisions at 5.02 TeV. We also derived and discussed the relation between the Tsallis temperature and the thermal temperature in TBW model which is the same as in the BGBW model.

\section*{Acknowledgements}
We would like to thank the ALICE collaboration for their share of the data in Xe-Xe collisions at 5.44 TeV. This work is supported by the Fundamental Research Funds for the Central Universities of China under GK201903022 and GK202003019, by the Scientific Research Foundation for the Returned Overseas Chinese Scholars, State Education Ministry, by Natural Science Basic Research Plan in Shaanxi Province of China (program No. 2020JM-289) and by the National Natural Science Foundation of China under Grant Nos. 11447024 and 11505108.

\section*{Appendix}
 \setcounter{equation}{0}
\setcounter{subsection}{0}
\renewcommand{\theequation}{A\arabic{equation}}
\renewcommand{\thesubsection}{A\arabic{subsection}}
We present the detailed derivation of Eq.(\ref{eq:eTBW}) for the TBW model and show how to obtain its slope at high $m_{T}$ or $p_{T}$.

In the TBW model, the invariant distribution function for identified particles is given by\cite{TBW_1}
\begin{equation}
	f(x, p)=\frac{g}{(2\pi)^{3}}\left(1+(q-1) \frac{E-\mu}{T}\right)^{-\frac{1}{q-1}},
\end{equation}
where $T $ is the Tsallis temperature, $g$ is the degeneracy factor, $E=p^{\nu} u_{\nu}$ represents the energy of emitted particles with momentum $p^{\nu}$ from the source moving with the velocity $u_{\nu}$, $p^{\nu}=(m_{\rm T} \cosh y, p_{\rm T} \cos \phi_{p}, p_{\rm T} \sin \phi_{p}, m_{\rm T} \sinh y)$, $u^{\mu}=\cosh \rho$ $\left(\cosh y_{s}, \tanh \rho \cos \phi_{b}, \tanh \rho \sin \phi_{b}, \sinh y_{s}\right)$,  the symbols $q$, $m_{\rm T}$,  $y$, $y_{s}$, $\phi_{p}$, $\phi_{b}$ and $\rho$ have been explained in sect. \ref{sec:method}. For the LHC energy regime, the chemical potential $\mu$ is set to be 0 due to the the near symmetry in particle-antiparticle production. The invariant momentum spectrum for identified particles then is written as 
\begin{equation}
	E \frac{d^{3} N}{d^{3} \boldsymbol{p}}=\frac{d^{3} N}{p_{\rm T} d p_{\rm T} d y d \phi_{p}}=\frac{g}{(2\pi)^{3}}\int_{\Sigma_{f}} \left[1+(q-1) \frac{p^{\nu} u_{\nu}}{T}\right]^{-\frac{1}{q-1}} p^{\lambda} d \sigma_{\lambda},
\end{equation}
where $\Sigma_{f}$ is the decoupling hypersurface, d$\sigma_{\lambda} $ is the normal vector to the hypersurface. With the parameterization of the surface in cylindrical coordinates\cite{hypersurface}, 
\begin{equation}
	d \sigma_{\lambda}=\left(r d \phi_{b} d r d z, -\boldsymbol{e_{r}} r d \phi_{b} d z d t, 0, -\boldsymbol{e_{z}} r d \phi_{b} d r d t\right),
\end{equation}
$p^{\lambda} d \sigma_{\lambda}$ can be read as 
\begin{equation}
p^{\lambda} d \sigma_{\lambda}=r d \phi_{b} d y_{s}\left[m_{\mathrm{T}} \tau \cosh (y_{s}-y) d r+m_{\mathrm{T}} \sinh (y_{s}-y) d \tau-p_{\mathrm{T}} \tau \cos \phi_{p} d \tau\right],
\end{equation}
where $y_{s}=\frac{1}{2}{\rm ln}\frac{t+z}{t-z}$, $\tau=\sqrt{t^2-z^2}$ is the longitudinal proper time. In the case of particle decoupling at $\tau_{0}$, the above equation is simplified as 
\begin{equation}
p^{\lambda} d \sigma_{\lambda}=\tau_{0}  m_{\mathrm{T}} \cosh (y_{s}-y) r d r d \phi_{b} d y_{s}.
\end{equation}
With the identity $p^{\nu} u_{\nu}=m_{\rm T} \cosh \rho \cosh (y_{s}-y)-p_{\rm T} \sinh \rho \cos \left(\phi_{p}-\phi_{b}\right)$, the identified particle spectrum then can be expressed as 
\begin{equation}\label{eq:spectrum}
  \small
  \begin{split}
\frac{d^{3} N}{p_{\rm T} d p_{\rm T} d y d \phi_{p}}&=\frac{g \tau_{0}}{(2 \pi)^{3}} \int_{\Sigma_{f}} d y_{s} r d r d \phi_{b} m_{T} \cosh (y_{s}-y)\\
 &\times \left[1+\frac{q-1}{T}\left[m_{\rm T} \cosh \rho \cosh (y_{s}-y)-p_{T} \sinh \rho \cos \left(\phi_{p}-\phi_{b}\right)\right]\right]^{-\frac{1}{q-1}} .
\end{split}
\end{equation}
With the integration over $d\phi_{p}$, in the mid-rapidity region with $y\approx 0$, the above equation becomes Eq. (\ref{eq:eTBW}).

As shown in Tables \ref{tab:pb_pb_2_76_TeV_fit_parameters}-\ref{tab:p_pb_5_02_TeV_fit_parameters}, the non-extensive parameter $q$ at the LHC energy regime,  ranging from 1.03 to 1.12, is close to unity. Thus we can perform a Taylor expansion for $f(x, p)$ at $\mu=0$ in a series of $q-1$ as follows\cite{Tsallis_dist_1}:
\begin{equation}
f(x, p)=e^{-\frac{E}{T}}+(q-1)\frac{1}{2}\left(\frac{E}{T}\right)^{2}e^{-\frac{E}{T}}+\mathcal{O}\left((q-1)^2\right).
\end{equation}
Applying this expansion to Eq. (\ref{eq:spectrum}) and neglecting the $\mathcal{O}\left((q-1)^2\right)$ terms, we can get the invariant identified particle spectrum as 
\begin{equation}\label{eq:spectrum_expand}
  \small
  \begin{split}
\frac{d^{3} N}{2\pi p_{\rm T} d p_{\rm T} d y }&=\frac{1}{2\pi}\int_{0}^{2\pi} \frac{d^{3} N}{p_{\rm T} d p_{\rm T} d y d \phi_{p}} d\phi_{p} \\
 &\approx \frac{g \tau_{0}}{(2 \pi)^{4}} \int_{0}^{2\pi} d\phi_{p}\int_{\Sigma_{f}} d y_{s} r d r d \phi_{b} m_{\rm T} \cosh (y_{s}-y)\\
 &\times \left[1+(q-1)\frac{1}{2} \left(\frac{m_{\rm T} \cosh \rho \cosh (y_{s}-y)-p_{\rm T} \sinh \rho \cos \left(\phi_{p}-\phi_{b}\right)}{T}\right)^2\right]\\
 &\times \exp \left[-\frac{m_{\rm T} \cosh \rho \cosh (y_{s}-y)-p_{\rm T} \sinh \rho \cos \left(\phi_{p}-\phi_{b}\right)}{T}\right].
\end{split}
\end{equation}
There are four terms when expanding the right-hand side of the above equation. The first term is exactly the BGBW distribution, 
\begin{equation}\label{eq:spectrum_expand_1}
\small
\frac{d^{3} N}{2\pi p_{\rm T} d p_{\rm T} d y }\bigg|_{\rm 1^{st}\ term}= A m_{\rm T} \int _{0}^{R} rdr d\phi_{b}K_{1}(\zeta_{m})I_{0}(\zeta_{p}),
\end{equation}
where $\zeta_{m}=m_{\rm T}\cosh \rho/T$, $\zeta_{p}=p_{\rm T}\sinh \rho/T$, $A=g \tau_{0}/(2 \pi)^{3}$. The second term refers to the integration of $\zeta_{m}^{2}\cosh^{2} (y_{s}-y)$ and is written as 
\begin{equation}\label{eq:spectrum_expand_2}
%\footnotesize
\small
\frac{d^{3} N}{2\pi p_{\rm T} d p_{\rm T} d y }\bigg|_{\rm 2^{nd}\ term}= AB m_{\rm T}^{3} \int _{0}^{R} rdr d\phi_{b}\left(\frac{K_{3}(\zeta_{m})}{4}+\frac{3K_{1}(\zeta_{m})}{4} \right)I_{0}(\zeta_{p})\cosh^{2} \rho,
\end{equation}
where $B=(q-1)/(2T^{2}) $.The third term corresponds to the integration of $\zeta_{p}^{2}\cos^{2} (\phi_{p}-\phi_{p})$ and is expressed as 
\begin{equation}\label{eq:spectrum_expand_3}
\small
\frac{d^{3} N}{2\pi p_{\rm T} d p_{\rm T} d y }\bigg|_{\rm 3^{rd}\ term}= A B m_{\rm T} p_{\rm T}^{2} \int _{0}^{R} rdr d\phi_{b}K_{1}(\zeta_{m})\left(\frac{I_{0}(\zeta_{p})}{2}+\frac{I_{2}(\zeta_{p})}{2} \right)\sinh^{2} \rho.
\end{equation}
The last term is the result of the integration on $\zeta_{m}\zeta_{p}\cosh(y_{s}-y)\cos(\phi_{p}-\phi_{b})$,
\begin{equation}\label{eq:spectrum_expand_4}
\small
\frac{d^{3} N}{2\pi p_{\rm T} d p_{\rm T} d y }\bigg|_{\rm 4^{th}\ term}=- A B m_{\rm T}^{2} p_{\rm T} \int _{0}^{R} rdr d\phi_{b} \left(K_{0}(\zeta_{m})+K_{2}(\zeta_{m}) \right)I_{1}(\zeta_{p}) \sinh \rho \cosh \rho.
\end{equation}

The slope of the TBW distribution in a semilogarithmic is 
\begin{equation}\label{eq:spectrum_slope}
\small
\begin{split}
\frac{d}{dm_{\rm T}} \ln \frac{d^{3} N}{2\pi p_{\rm T} d p_{\rm T} d y } &= \frac{d}{dm_{\rm T}}\ln \Bigg\{ m_{T} K_{1}(\xi_{m}) I_{0}(\xi_{p})\\
&+B \bigg[\frac{m_{\rm T}^{3}\cosh^{2} \rho}{4} \left(K_{3}(\zeta_{m})+3K_{1}(\zeta_{m}) \right)I_{0}(\zeta_{p}) \\
&+  \frac{m_{\rm T} p_{\rm T}^{2}\sinh^{2} \rho}{2} K_{1}(\zeta_{m})\left(I_{0}(\zeta_{p})+I_{2}(\zeta_{p}) \right)\\
&- m_{\rm T}^{2} p_{\rm T} \sinh \rho \cosh \rho \left(K_{0}(\zeta_{m})+K_{2}(\zeta_{m}) \right)I_{1}(\zeta_{p}) \bigg] \Bigg\}.
\end{split}
\end{equation}
In the case of $m_{\rm T} \sim p_{\rm T} \gg T$, $m_{\rm T} / p_{\rm T} \to 1$, $I_{1,2,3}/I_{0} \to 1$, $K_{0,2,3,4}/K_{1} \to 1$, the first term of the right-hand side in the above equation is 
\begin{equation}\label{eq:spectrum_slope_1}
\small
\lim_{m_{\rm T}\to \infty}\frac{d}{dm_{\rm T}} \ln \frac{d^{3} N}{2\pi p_{\rm T} d p_{\rm T} d y } \bigg|_{\rm 1^{st}\ term} =0.
\end{equation}
The second term reads as 
\begin{equation}\label{eq:spectrum_slope_2}
\small
\lim_{m_{\rm T}\to \infty}\frac{d}{dm_{\rm T}} \ln \frac{d^{3} N}{2\pi p_{\rm T} d p_{\rm T} d y } \bigg|_{\rm 2^{nd}\ term} =\frac{\cosh^2 \rho (\sinh \rho - \cosh \rho)}{T(\cosh^2 \rho+\sinh^2 \rho-2\cosh \rho \sinh \rho) }.
\end{equation}
The third term becomes 
\begin{equation}\label{eq:spectrum_slope_3}
\small
\lim_{m_{\rm T}\to \infty}\frac{d}{dm_{\rm T}} \ln \frac{d^{3} N}{2\pi p_{\rm T} d p_{\rm T} d y } \bigg|_{\rm 3^{rd}\ term} =\frac{\sinh^2 \rho (\sinh \rho - \cosh \rho)}{T(\cosh^2 \rho+\sinh^2 \rho-2\cosh \rho \sinh \rho) }.
\end{equation}
The last term is written as 
\begin{equation}\label{eq:spectrum_slope_4}
\small
\lim_{m_{\rm T}\to \infty}\frac{d}{dm_{\rm T}} \ln \frac{d^{3} N}{2\pi p_{\rm T} d p_{\rm T} d y } \bigg|_{\rm 4^{rth}\ term} =\frac{-2\cosh \rho \sinh \rho (\sinh \rho - \cosh \rho)}{T(\cosh^2 \rho+\sinh^2 \rho-2\cosh \rho \sinh \rho) }.
\end{equation}
Combining the above four terms together, we can get 
\begin{equation}\label{eq:spectrum_slope_tot}
\small
\lim_{m_{\rm T}\to \infty}\frac{d}{dm_{\rm T}} \ln \frac{d^{3} N}{2\pi p_{\rm T} d p_{\rm T} d y }  =\frac{\sinh \rho - \cosh \rho}{T}=-\frac{1}{T}\sqrt{\frac{1-\langle \beta \rangle}{1+\langle \beta \rangle}}.
\end{equation}
Thus the thermal temperature in the TBW model is 
\begin{equation}\label{eq:spectrum_slope_tot}
\small
T_{th}=T\sqrt{\frac{1+\langle \beta \rangle}{1-\langle \beta \rangle}},
\end{equation}
which is similar to that in the BGBW model\cite{BGBW_dist}.

\end{document}